\documentclass[a4paper,11pt]{article}
\pdfoutput=1 % if your are submitting a pdflatex (i.e. if you have
\usepackage{jcappub} % for details on the use of the package, please
\usepackage[T1]{fontenc}
\usepackage{ae,aecompl}
\usepackage{graphicx}	% Including figure files
\usepackage{amsmath}	% Advanced maths commands
\usepackage{amssymb}	% Extra maths symbols
\usepackage{lipsum}
\usepackage{mathtools}
\usepackage{comment}
\usepackage{multirow}
\usepackage{xcolor}
\usepackage{hyperref}
\usepackage{latexsym}
\usepackage{pifont}

\graphicspath{{./Pictures/}}

\newcommand{\tvec}[1]{\mathbf{#1}}      % vector
\newcommand{\de}{\,\mathrm{d}}          % differential
\newcommand{\der}[2]{\frac{\mathrm{d}#1}{\mathrm{d}#2}} % total derivative
\newcommand{\derp}[2]{\frac{\partial #1}{\partial #2}}  % partial derivative
\newcommand{\xmark}{\ding{55}}          % X

\newcommand{\Neff}{\ensuremath{N_{\rm eff}}}
\title{\boldmath Neutrino clustering in the Milky Way and beyond}

\author[a]{P.\ Mertsch,}
\emailAdd{pmertsch@physik.rwth-aachen.de}

\author[b,c,d]{G.\ Parimbelli,}
\emailAdd{gparimbe@sissa.it}

\author[e]{P.F.\ de Salas,}
\emailAdd{pablo.fernandez@fysik.su.se}

\author[f]{S.\ Gariazzo,}
\emailAdd{gariazzo@ific.uv.es}

\author[a]{J.\ Lesgourgues}
\emailAdd{lesgourg@physik.rwth-aachen.de}

\author[f]{and S.\ Pastor}
\emailAdd{pastor@ific.uv.es}

\affiliation[a]{Institute for Theoretical Particle Physics and Cosmology (TTK)\\
RWTH Aachen University, D-52056 Aachen, Germany}

\affiliation[b]{SISSA - International School for Advanced Studies, Via Bonomea 265, 34136 Trieste, Italy}

\affiliation[c]{INFN - National Institute for Nuclear Physics, Via Valerio 2, 34127 Trieste, Italy}

\affiliation[d]{IFPU - Institute for Fundamental Physics of the Universe, Via Beirut 2, 34151 Trieste, Italy}

\affiliation[e]{The Oskar Klein Centre for Cosmoparticle Physics,
Department of Physics, Stockholm University, SE-106 91 Stockholm, Sweden}

\affiliation[f]{Instituto de F\'{\i}sica Corpuscular
(CSIC-Universitat de Val\`{e}ncia)\\
Parc Cient\'{\i}fic UV, C/ Catedr\'atico Jos\'e Beltr\'an, 2, E-46980 Paterna (Valencia), Spain}

\abstract{
The standard cosmological model predicts the existence of a Cosmic Neutrino Background, which has not yet been observed directly.
Some experiments aiming at its detection are currently under development, despite the tiny kinetic energy of the cosmological relic neutrinos, which makes this task incredibly challenging.
Since massive neutrinos are attracted by the gravitational potential of our Galaxy, they can cluster locally.
Neutrinos should be more abundant at the Earth position than at an average point in the Universe.
This fact may enhance the expected event rate in any future experiment.
Past calculations of the local neutrino clustering factor only considered a spherical distribution of matter in the Milky Way and neglected the influence of other nearby objects like the Virgo cluster, although recent $N$-body simulations suggest that the latter may actually be important.
In this paper, we adopt a back-tracking technique, well established in the calculation of cosmic rays fluxes,
to perform the first three-dimensional calculation of the number density of relic neutrinos at the Solar System, taking into account not only the matter composition of the Milky Way,
but also the contribution of the Andromeda galaxy and the Virgo cluster.
The effect of Virgo is indeed found to be relevant and to depend non-trivially on the value of the neutrino mass.
Our results show that the local neutrino density is enhanced by 0.53\% for a neutrino mass of $10$~meV, 12\% for $50$~meV, $50\%$ for $100$~meV or 500\% for $300$~meV.
}

\begin{document}

\hfill{TTK-19-44}

\maketitle
\flushbottom

%======================================================================
%======================================================================
%======================================================================
%========================= PAPER STARTS HERE ==========================
%======================================================================
%======================================================================
%======================================================================

\section{Introduction}
\label{sec:intro}

The existence of a background of relic neutrinos,
produced in the early Universe in a similar way as the photons that constitute the Cosmic Microwave Background (CMB),
is one of the yet unconfirmed predictions of the standard model of cosmology.
While we have good indirect indications,
such as the number of relativistic species in the early Universe,
which is constrained \cite{Aghanim:2018eyx}
to be very close to the expected theoretical value $\Neff=3.045$ \cite{Mangano:2005cc,deSalas:2016ztq,Gariazzo:2019gyi},
or the imprint of relativistic species on the CMB spectrum,
which is compatible with those of free-streaming relics (see e.g.\ \cite{Audren:2014lsa}),
a direct probe of the existence of a background of relic neutrinos would be a major discovery
and a confirmation of what we know about cosmology and neutrinos.
In particular, it would discard the possibility that ordinary neutrinos have decayed at some stage during the evolution of the universe
(see e.g.\ \cite{Beacom:2004yd,Chacko:2019nej})
or have been produced with unexpectedly low abundance
(e.g.\ in low reheating scenarios \cite{deSalas:2015glj}),
while another form of dark radiation would contribute to $\Neff\simeq3$.

The most promising method for obtaining a direct detection of the relic neutrino background
is to exploit neutrino capture on $\beta$-decaying nuclei \cite{Weinberg:1962zza},
in particular tritium \cite{Cocco:2007za}.
Using this method one would look for a (small) peak in the electron energy spectrum of tritium
due to the capture of relic neutrinos, just above the endpoint of $\beta$ decay.
Although such determination is a real challenge due to the required energy resolution
(that must be comparable with the absolute value of the neutrino mass)
and the high number of background events, coming from $\beta$ decay, that the experiment has to distinguish
from the signal,
a project named PTOLEMY \cite{Baracchini:2018wwj} is nowadays starting to test innovative technology
that could lead, for favorable values of the neutrino masses,
to the first direct observation of relic neutrinos \cite{Betti:2019ouf}.
The successful observation of relic neutrinos by PTOLEMY would also offer the unique opportunity to study for the first time
the interactions of non-relativistic neutrinos%
\footnote{Given the values of the mass splittings provided by neutrino oscillation experiments,
the second-to-lightest neutrino mass eigenstate must be heavier than at least 8~meV,
see e.g.\ \cite{Capozzi:2018ubv,deSalas:2017kay,Esteban:2018azc}, while the mean energy of relic neutrinos is of the order of $10^{-4}$ eV.}.

Since the number of events that a direct detection experiment could measure
depends linearly on the local number density of relic neutrinos,
it is important to have a precise knowledge of how many relic neutrinos are present today at Earth.
The average number density of neutrinos predicted by the standard cosmological model
is 56~cm$^{-3}$ per family and per degree of freedom.
Relic neutrinos, however, have a very small mean energy today (of the order of $10^{-4}$~eV),
therefore
their average number density can be enhanced because of the gravitational attraction of the matter content of the Galaxy,
as well as other neighbouring galaxies and galaxy clusters, provided that their masses are large enough
to let them cluster at small scales.
The calculation of the clustered number density of relic neutrinos was proposed for the first time
in \cite{Singh:2002de} using a method based on the collisionless Boltzmann equation,
and in \cite{Ringwald:2004np} using a method called $N$-one-body simulations.
The latter case consists in computing the trajectories of several ($N$) independent test particles (one-body)
in the evolving gravitational potential of the Galaxy,
starting from some high redshift until today,
and then reconstructing the profile of the neutrino halo
according to the final positions of all the test particles.
The same method has been adopted later in \cite{deSalas:2017wtt,Zhang:2017ljh},
where an updated treatment of the dark matter (DM) and baryonic content of the Milky Way was considered.

In this paper, we improve the calculation presented in \cite{deSalas:2017wtt}.
For the first time, we take into account not only the contribution of the Milky Way, but also
the contribution of relevant, nearby astrophysical objects, such as the Andromeda galaxy and the Virgo cluster.
In order to perform this task, we need to relax the assumption of spherical symmetry
that has been used in previous works.
In sections \ref{sec:theory} to \ref{sec:code} we discuss the theoretical aspects of the calculation
and the practical implementations in our code.
The treatment of the matter content of the two galaxies and the Virgo cluster
is presented in section \ref{sec:gravitational_potential}.
In the two final sections \ref{sec:results} and \ref{sec:conclusions}
we present and discuss our results on the local number density of relic neutrinos.

%======================================================================
%======================================================================
\section{Theory}
\label{sec:theory}

The motion of the $N$ test particles must be computed in an evolving matter background.
On galaxy scales and at recent times, the behavior of at least the two heaviest neutrino states is well captured by Newton's theory, in which the
motion equations can be obtained from the following Lagrangian:
\begin{equation}
\mathcal{L}
=a\left(\frac{1}{2}m_\nu v^2-m_\nu\Phi(\vec x, t)\right)\,,
\end{equation}
where $a$ is the scale factor,
$m_\nu$ is the mass of the test neutrino,
$v$ its velocity
and $\Phi$ the gravitational potential.
Let us now write the corresponding Hamiltonian,
expressed in Cartesian coordinates $x,y,z$ and the corresponding conjugate momenta $p_x, p_y, p_z$:
\begin{equation}
\mathcal{H}
=
\frac{1}{2a m_\nu}\left(p_x^2+p_y^2+p_z^2\right)+a m_\nu\Phi(\vec x, t)\,.
\label{eqn:Hamiltonian}
\end{equation}
We can now compute the equations of motion.
Denoting with a dot the derivative with respect to conformal time, we find
\begin{equation}
p_i=a m_\nu\dot x_i,
\qquad
\dot p_i=-am_\nu\derp{\Phi(\vec x,t)}{x^i},\quad\text{with }x_i={x,y,z}\,.
\label{eqn:eoms1}
\end{equation}
For a spherically symmetric potential, the equations of motion simplify significantly due to the conservation of angular momentum and are best expressed in spherical coordinates.
Note that previous works always considered a spherically symmetric Milky Way.

The most efficient way to do the calculations, however,
does actually not involve the solution of the above equations.
It is more convenient to rescale the momenta of the test particles
in order to eliminate the neutrino mass from the equations:
\begin{equation}
u_i = p_i/m_\nu\,,
\end{equation}
thus replacing $p_i\rightarrow u_i$ and $m_\nu\rightarrow1$ in the Hamilton equations above.
Solving for the velocity $u_i$ using a single neutrino mass will allow to obtain the results
for different neutrino masses,
simply rescaling the parameter space volume appropriately (see \cite{Ringwald:2004np,Zhang:2017ljh}).

In order to solve the equations of motion, we need the gravitational potential $\Phi$
of the Galaxy as well as those of other nearby objects like Andromeda and Virgo.
We will make use of the Poisson equation to obtain the contribution to the total gravitational potential of each component described by its energy density $\rho$,
\begin{equation}
4\pi G a^2 \rho
=
\nabla^2\Phi(\vec x, t)\label{eq:poisson_lapl}\,,
\end{equation}
where the Laplacian operator is in comoving coordinates.
Since the Poisson equation is linear, it is always possible to solve it separately for the different constituents of the total matter density.
When assuming spherical symmetry, the potential depends only on the distance from the center of the halo, $r$,
so that it is possible to have an analytic expression for the derivative of the potential
from equation \eqref{eq:poisson_lapl}:
\begin{equation}
\derp{\Phi}{r}(r, z)
=
\frac{4\pi G a^2}{r^2}\int_0^r
\rho_{\rm halo}(x, z)\,x^2\, \de x
=
\frac{G}{ar^2}M_{\rm halo}(r, z)\,.
\label{eq:dphidr}
\end{equation}
When the density is instead not symmetric, the Poisson equation must be solved numerically.
The most convenient way is to use Fourier transforms,
as we discuss in the appendix \ref{app:poisson}.
Once $\Phi$ is obtained, one has to compute the partial derivatives
that enter in the Hamilton equations above and that we discuss in details in section~\ref{sec:gravitational_potential}.

The $N$-one-body simulation method requires the solution of the equations of motion of many test particles
with different initial conditions.
When dealing with the spherically symmetric case,
one has to sample different values for the parameter space of only three quantities:
the initial distance from the center of the halo,
the initial momentum of the particle,
and the initial angle between the initial momentum vector and the radial direction.
Because of spherical symmetry, the motion of each test particle will always be contained in a plane.
The calculation of the number density profile of the relic neutrino halo (and of its particular value at Earth)
takes into account the final position of all test particles
weighted by their
initial phase space, see \cite{Ringwald:2004np}.
If one wants to relax the assumption of spherical symmetry, however, the final position will have to be computed
as a function of six input variables (three for the position and three for the momentum) instead of just three.
Since the number of test particles that are required in order to obtain a sufficiently precise result
scales exponentially with the number of dimensions,
repeating the calculation
without spherical symmetry would require an unreasonably high number of computational hours.
Moreover, many of the simulated test particles will end up very far from the position of the Earth,
and will
give very little or no contribution to the local density of relic neutrinos.

Fortunately, a simple way to overcome this problem has been known for many years in the context of cosmic ray propagation.
Instead of forward-tracking the particles starting from homogeneous and isotropic initial conditions at high redshift,
it is more efficient to consider only those particles that are at Earth today.
This is done by inverting the arrow of time in the equations, and back-tracking the particles
from our position today.
Afterwards, we can attribute an initial phase-space volume and an appropriate statistical weight to each trajectory.
The main advantage of this method is that one only needs to sample over the
3-momentum of the neutrinos reaching the Earth today, since their position is fixed by assumption,
regardless of the assumed symmetries of the astrophysical environment.
The computational time will thus remain comparable to that of previous works assuming spherical symmetry, while allowing us to introduce a complex distribution of matter with many objects.

The drawback, however, is that with the back-tracking method
one does not obtain the full shape of the neutrino halo around the Earth,
but only
the local number density.
To estimate the shape of the neutrino profile, multiple simulations at different positions are required.
More details on the back-tracking method and on our specific implementation are discussed in the next sections.

%======================================================================
%======================================================================
\section{Forward versus backward $N$-one-body method}

Some of us used the forward-tracking technique in a previous publication \cite{deSalas:2017wtt}, following Appendix A.3 in \cite{Ringwald:2004np} and using the kernel method of \cite{Merritt:1994}.
In this approach,
the number density is reconstructed from a set of $N$ representative particles of the phase-space interval $(r_a,p_{r,a},p_{T,a})_i \rightarrow (r_b,p_{r,b},p_{T,b})_i$.
Each trajectory is given a
weight $w_i$ $(i=1,\cdots , N)$
\begin{equation}
    w_i = \int_{(r_a,p_{r,a},p_{T,a})_i}^{(r_b,p_{r,b},p_{T,b})_i} \int_{\theta , \phi , \varphi} f(p) \ \de^3 r \de^3 p,
    \label{weight}
\end{equation}
where $f(p)$ is assumed to be the homogeneous and isotropic Fermi-Dirac distribution (neglecting small linear perturbations far from the Milky Way),
while we used $\de^3 r = r^2 \sin\theta \de\theta \de\phi \de r$ and $\de^3 p = p_T \de p_T \de p_r \de\varphi$,
$p_T$ being the transverse momentum and $p_r$ the radial momentum.
The final number density at radius $r$ is then given by
\begin{equation}
    n(r) = \sum_{i=1}^{N} \frac{w_i}{\xi^3} K(r,r_i,\xi),
\end{equation}
where the Gaussian kernel
\begin{equation}
    K(r,r_i, \xi) = \frac{\xi^2}{(2\pi)^{3/2} r r_i} \exp \left( - \frac{r^2 + r_i^2}{2\xi^2} \right) \sinh \left( \frac{rr_i}{\xi^2}\right)
\end{equation}
plays the role of smearing the particles around the surface of a sphere in order to get a profile that is spherically symmetric.
The parameter $\xi$ is the window width \cite{Merritt:1994} and its value can be optimized for each step in the simulation.

When switching to the back-tracking method, we take the opposite perspective.
We draw trajectories from samples of the neutrino momentum today, at the position of the Earth.
At that time and location, the phase-space distribution of neutrinos is no longer close to the Fermi-Dirac distribution of the average neutrino background, due to the non-linear dynamics inside the halo.
Fortunately, we can make use of Liouville's theorem to compute the statistical weight of each phase-space volume element around the Earth to that at the other end of the trajectory,
where neutrinos still obey the average homogeneous and isotropic Fermi-Dirac distribution.

Liouville's theorem~\cite{2002clme.book.....G} implies the conservation of phase-space density along the solutions of the equations of motions,
\begin{equation}
\dot{f} + \dot{\vec{x}} \cdot \vec{\nabla}_{\vec{x}} f + \dot{\vec{p}} \cdot \vec{\nabla}_{\vec{p}} f = 0 \, .
\end{equation}
After tracking a particle from redshift $z=0$, the Earth position $\vec{x}_{\oplus}$ and some arbitrary momentum $\vec{p}_j(0)$
back to redshift $z_{\text{back}}$, position $\vec{x}_j(z_{\text{back}})$ and momentum $\vec{p}_j(z_{\text{back}})$,
we can compute the phase-space distribution today and in the right direction by applying
\begin{equation}
f(\vec{x}_{\oplus}, \vec{p}_j(0), 0) = f_\mathrm{back}(\vec{x}_j(z_{\text{back}}), \vec{p}_j(z_{\text{back}}), z_{\text{back}}) \, ,
\end{equation}
where $f_\mathrm{back}$ can be identified with the Fermi-Dirac distribution of the average neutrino background.
This gives us $f$ today in any direction.
The final number density is then obtained by integrating over the observed momentum $\vec{p}_j(0)$, without any need for Gaussian smoothing.
In this work we have chosen $z_{\text{back}} = 4$,
but we verified that the value of $z_{\text{back}}$ has no significant impact
on the final number density, see section~\ref{sec:results}.
Note that this Liouville mapping is routinely used when back-tracking cosmic rays, see e.g.\ Ref.~\cite{Mertsch:2019elt}.

In both methods, after obtaining the local number density $n_{\nu_i}(\vec{x}_{\oplus})$ for each mass eigenstate $i$,
one can compute the clustering factor:
\begin{equation}
f_i
\equiv
n_{\nu_i}/n_{\nu,0}
\,,
\end{equation}
where
$n_{\nu,0}=112 \ \mathrm{cm}^{-3}$ is the cosmological average number density for one family of neutrinos plus anti-neutrinos.

%======================================================================
%======================================================================
\section{Computing neutrino clustering with back-tracking}
\label{sec:code}

For solving the equations of motion, we
have used a symplectic ODE solver that also conserves phase-space volume,
the \texttt{symplectic\_rkn\_sb3a\_mclachlan} solver, that is the symmetric B3A method of the Runge-Kutta-Nystr\"om scheme of sixth order~\cite{McLachlan:1995}
from the \texttt{odeint} package of the \texttt{Boost} libraries\footnote{\url{http://www.boost.org}}~\cite{Schaeling:2014}.

The symplectic solvers of the \texttt{odeint} package require the equations of motion to be separable,
that is the time-derivatives of the coordinates are functions of the conjugate momenta only and \textit{vice versa}, and autonomous,
that is all right-hand sides must not depend on time $t$ explicitly.
The latter requirement poses a problem since both the expanding space-time and the redshift evolution of the gravitational potential
introduce a time dependence in the Hamiltonian, cf.\ eq.~\eqref{eqn:Hamiltonian}.
A common fix consists of treating time as an extra variable to be integrated on top of $u_i(t)$ and $x_i(t)$, with a trivial derivative $\dot{t}=1$ (cf.\ e.g.~\cite{Blanes:2001}).
With such an addition, the system is formally autonomous and still separable.
Finally, we note that if we substitute time for the new variable $s$,
\begin{equation}
s(z) = -\int_0^z \frac{\de z}{\dot{a}} \, ,
\end{equation}
the equations of motion \eqref{eqn:eoms1} take on the even simpler form
\begin{equation}
\der{x_i}{s} = u_i \, , \quad \der{u_i}{s} = - a^2 \derp{\phi}{x_i} \, ,
\label{eqn:eoms2}
\end{equation}
allowing to further
speed up the computation.

Although back-tracking dramatically reduces the number of particles to be simulated, we still have to deal with a large number of trajectories, obtained by solving
eqs.~\eqref{eqn:eoms2}
for several initial conditions $u_i(z=0)$. This requirement is most efficiently fulfilled in a ``Single Instruction, Multiple Data'' (SIMD) architecture,
modern graphic processing units (GPUs) being an example.
We have used the \texttt{CUDA} framework\footnote{\url{http://developer.nvidia.com/cuda-zone}} via the \texttt{Thrust} library\footnote{\url{http://thrust.github.io}} which can be interfaced with \texttt{odeint}'s solvers.
In order to increase speed, we have pre-computed the baryonic contributions to the gravitational potential (see next section) at redshift $z=0$ on a grid in cylindrical coordinates $R$ and $z$,
loaded them as textures onto the GPU, bi-linearly interpolated them between grid points, and finally scaled the results up to higher redshifts $z$.
For the results below, we have isotropically sampled the arrival directions of neutrinos (20 points for polar angle, 20 points for azimuth) and logarithmically sampled in momentum (100 points over 3 decades), which leads to a grid of $4 \times 10^4$ velocities.
We have checked that this is sufficient for getting well-converged clustering factors even in the non-axisymmetric case (Milky Way dark matter plus baryons plus Andromeda and Virgo).
All the computations were performed on an Nvidia Quadro P6000.
Depending on the number of different contributions to the gravitational potential, back-tracking the $4 \times 10^4$ particles from redshift $z=0$ to $z=4$ required between $120$ and $500 \, \text{minutes}$.

%======================================================================
%======================================================================
\section{Density profiles and gravitational potential}
\label{sec:gravitational_potential}

In this Section we describe how we implement the gravitational potential of the objects that we include in our analysis.
For the Milky Way, we consider a spherical dark matter halo plus a number of baryonic components,
which are described in axial symmetry.
Beyond the Milky Way, we consider spherical dark matter halos for the Andromeda galaxy and the Virgo Cluster,
which are the largest objects relatively close to Earth
that can have an impact on the local density of relic neutrinos.
Finally, we report technical details on the discretization of the grid that we adopt in the numerical calculation
for the interpolation of the derivatives.

%======================================================================
\subsection{The Milky Way}
\label{sec:mw_potential}

For the dark matter in our Galaxy, we consider two distinct cases: a Navarro-Frenk-White (NFW, \cite{Navarro:1995iw}) and an Einasto \cite{Einasto:1965czb} profile, which read, respectively,
\begin{align}
\rho_\textrm{NFW}(r)
&=
\cfrac{\rho_0}{\displaystyle\frac{r}{R_s}\left(1+\frac{r}{R_s}\right)^2}
\quad \textrm{for } r<R_\mathrm{vir},
\label{eq:nfw_profile}
\\
\rho_\textrm{Ein}(r)
&=
\rho_0 \exp[-(r/R_s)^\alpha],
\label{eq:einasto_profile}
\end{align}
where $\rho_0$ is the normalization, $R_s$ is the scale radius, $R_\mathrm{vir}$ is the virial radius of the NFW profile (which is related to $R_s$ through the concentration parameter $c=R_\mathrm{vir}/R_s$),
and $\alpha$ is an additional parameter for the Einasto profile that controls the change in the slope of the density.

Concerning baryons, we follow the treatment of \cite{Misiriotis+06}
and adopt five separate components: stars, warm and cold dust, atomic HI and molecular H$_2$ gas.
The density of stars is parameterized using a disk plus a bulge.
The bulge of the Milky Way has been shown to have a triaxial shape \cite{Portail+15}.
However, since we are mainly interested in the neutrino clustering at the Earth position,
which is located at distances ($8.2\pm0.1$~kpc \cite{Bland-Hawthorn:2016})
significantly larger than the bulge size,
we can safely approximate it as a sphere.
In particular, again following \cite{Misiriotis+06},
we assume for the bulge profile a Sersic law with index $n=4$ (i.e.\ a de Vaucouleurs profile):
\begin{equation}
\rho_\mathrm{DeVac}(r) = \rho_0 \ \exp\left[-A\left(\frac{r}{R_b}\right)^{1/4}\right] \ \left(\frac{r}{R_b}\right)^{-7/8},
\label{eq:bulge_rho}
\end{equation}
where $A=2n-1/3\approx 7.67$ and we take $R_b = 0.74\,\mathrm{kpc}$.
The other baryonic components
are assumed to be distributed
according to a double exponential disk profile
\begin{equation}
    \rho_{\exp}(R,z) = \rho_0 \ e^{-R/R_s} \ e^{-|z|/z_s}.
    \label{eq:exponential_disk}
\end{equation}

The present day values of the parameters of the different profiles
are obtained as follows.
The parameters related to the dark matter component of the Milky Way at $z=0$
are obtained by fitting the dark matter contribution to the rotation curve data as reported in \cite{Pato:2015tja},
following the same procedure already adopted in \cite{deSalas:2017wtt}%
\footnote{Notice that to switch from our parameterization of the Einasto profile in Eq.~\eqref{eq:einasto_profile} to the one used by \cite{deSalas:2017wtt},
one has to substitute
$\rho_0\rightarrow \rho_0 \exp{(-2/\alpha)}$ and
$R_s\rightarrow R_s \left(2/\alpha\right)^{1/\alpha}$.}.
Better estimates of the Galactic rotation curve are nowadays accessible thanks to the second data release of the ESA/Gaia mission \cite{Brown:2018dum}
(see e.g.\ \cite{deSalas:2019pee} for an analysis of the dark matter contribution to the rotation curve data presented in \cite{Eilers+18}).
However, given the limited radial extent of the data,
instead of fixing the total dark matter mass to the values predicted either in \cite{Pato:2015tja} or \cite{deSalas:2019pee},
we use an estimate based on orbiting Milky Way satellites up to $\sim 300\,\mathrm{kpc}$ from the center of the Galaxy \cite{Watkins:2010}.
Concerning the baryon components, we take the warm dust, cold dust, H$_2$ and HI profile parameters from \cite{Misiriotis+06},
as well as the scale parameters of the bulge and the disk.
For the central value of the density of the bulge we use \cite{Bland-Hawthorn:2016},
which also provides an estimate for the total stellar mass in the Galaxy ($5\times10^{10}\;M_\odot$).
From this number we can derive the total mass of the stellar disk by subtracting the total mass of the bulge.
The values of the parameters that we adopt are listed in Tables \ref{tab:dm_profile_parameters_z0} and \ref{tab:baryon_profile_parameters_z0}.

Regarding the HI density profile,
observations (e.g.\ \cite{Kalberla+05, McMillan+17}) have shown that the distribution of neutral hydrogen
in the outskirts of the Galaxy follows an exponential profile, as we assume in this work;
conversely, the central 2.75~kpc \cite{McMillan+17} seem to be devoid of it.
This feature would in principle spoil the analyticity of our potentials.
However, we found that neglecting the central hole in the hydrogen distribution,
i.e.\ extrapolating the exponential profile until the origin of the coordinates,
would just cause an increase of the total HI mass of 1\%,
which is in turn an overestimate of order 0.01\% on the total mass of the Milky Way.
We then feel safe to ignore such a feature in the HI profile, and we consider it to be also a double exponential disk, following eq.~\eqref{eq:exponential_disk}.

\begin{table}
\centering
\begin{tabular}{c|cc}
  & \textbf{NFW} & \textbf{Einasto} \\
  \hline
  \hline
  $M_\mathrm{vir}$ [$M_\odot$] & $2.03\times 10^{12}$ & $1.17\times 10^{12}$ \\
  \hline
  $\rho_0$ [$M_\odot$/kpc$^3$] & $1.06\times 10^7$    & $2.70\times 10^8$ \\
  \hline
  $R_s$ [kpc]                  & 19.9                 & 0.737 \\
  \hline
  $R_\mathrm{vir}$ [kpc]       & 333.5                & \color{red}{\xmark} \\
  \hline
  $\alpha$                     & \color{red}{\xmark}               & 0.45 \\
  \hline
\end{tabular}
\caption{Dark matter density parameters for the Milky Way at $z=0$,
obtained by fitting the data from \protect\cite{Pato:2015tja},
following the same procedure as in \protect\cite{deSalas:2017wtt}.}
\label{tab:dm_profile_parameters_z0}
\end{table}

\begin{table}
\centering
\begin{tabular}{c|cccc}
  & $\rho_0$ [$M_\odot$/kpc$^3$] & $R_s$ [kpc] & $z_s$ [kpc] & $M_\mathrm{tot}$ [$M_\odot$] \\
  \hline
  \hline
  Bulge       & $1.79\times10^{12}$
              & 0.74
              & \color{red}{\xmark}
              & $1.55\times10^{10}$
              \\
  \hline
  Disk        & $3.40\times 10^9$
              & 2.4
              & 0.14
              & $3.45\times 10^{10}$
              \\
  \hline
  Warm dust   & $1.80\times 10^4$    & 3.3   & 0.09                    & $2.22\times 10^5$    \\
  \hline
  Cold dust   & $2.23\times 10^6$    & 5.0   & 0.1                     & $7.01\times 10^7$    \\
  \hline
  H$_2$       & $2.00\times 10^8$    & 2.57  & 0.08                    & $1.33\times 10^9$    \\
  \hline
  HI          & $7.90\times 10^6$    & 18.24 & 0.52                    & $1.72\times 10^{10}$ \\
  \hline
\end{tabular}
\caption{Density profile parameters for the baryonic components at $z=0$.
We also provide the total mass for each component.
All the components have a profile described by equation \eqref{eq:exponential_disk},
except for the bulge, which follows a de Vaucouleurs profile (equation \eqref{eq:bulge_rho}).
The scale radii and heights are taken from \protect\cite{Misiriotis+06}, as specified in the main text.
The redshift evolution of the total mass is found following
the $N$-body simulation results of \protect\cite{Marinacci+13},
while we assume that $R_s$ and $z_s$ do not evolve in time.
}
\label{tab:baryon_profile_parameters_z0}
\end{table}

In order to compute the clustering factor today,
we also need the time evolution of the density profiles.
As we comment in section~\ref{sec:results},
most of the clustering happens at small redshifts,
so there is no need to compute the density profiles very precisely at all times.
The evolution in redshift of the density profiles is accounted for in the following way.
We assume the total virial mass of the dark matter halo to be constant in redshift,
while the concentration parameter changes according to \cite{Dutton:2014xda}
\begin{equation}
    \log \beta c_\mathrm{vir}(z) = a(z) + b(z) \log\left(\frac{M_\mathrm{vir}}{10^{12} \ h^{-1} \ M_\odot}\right),
    \label{eq:concentration}
\end{equation}
where
$\beta$ is a parameter (considered constant in time) which denotes the offset
of the Milky Way concentration with respect to the average one.
The functions $a(z)$ and $b(z)$ are different for the NFW and Einasto profiles. For the NFW they correspond to
$a(z) = 0.537 + (1.025 - 0.537) \exp\left[-0.718 \ z^{1.08}\right]$,
$b(z) = -0.097 +0.024 \ z$,
while for the Einasto profile
$a(z) = 0.459 + (0.977 - 0.459) \exp\left[-0.490 \ z^{1.303}\right]$,
$b(z) = -0.130 +0.029 \ z$.

The evolution of the virial radius is obtained from
\begin{alignat}{1}
M_{\rm vir}
&=
4\pi a^3\int_0^{R_{\rm vir}(z)}
\rho(x, z)\, x^2\, \de x,
\label{eq:virial_mass}
\\
R_\mathrm{vir}(z)
&=
\left(\frac{3 M_\mathrm{vir}}{4\pi \Delta_\mathrm{vir}(z) \rho_\mathrm{crit}(z)}\right)^{1/3},
\label{eq:virial_radius}
\end{alignat}
where $\rho_\mathrm{crit}=3H_0^2/(8\pi G)$ is the critical density of the Universe and
$\Delta_\mathrm{vir}(z) = 18\pi^2 + 82\left[\Omega_m(z)-1\right] - 39\left[\Omega_m(z)-1\right]^2$ \cite{Bryan+98} for the NFW.
For the Einasto profile it is instead fixed to $\Delta_\mathrm{vir} = 200$, since this was the approach followed by \cite{Dutton:2014xda} in order to obtain the numerical values of the corresponding $a(z)$ and $b(z)$ equations.
Combining equations \eqref{eq:concentration}, \eqref{eq:virial_mass} and \eqref{eq:virial_radius}
allows us to find the scale radius as a function of redshift.
The cosmology used in this work has $h = 0.6766$ and $\Omega_m = 0.3111$ according to the Planck (TT,TE,EE+lowE+lensing+BAO) best-fit model \cite{Aghanim:2018eyx}.

On the other hand, reconstructing the evolution of scale radii of baryon components is a hard task.
For simplicity, we assume that the radii are constant in time, while the central densities change according
to the results of $N$-body simulations obtained by \cite{Marinacci+13}.
In particular we assume that the fraction of each component with respect to the total baryon mass is conserved.

For the equations of motion, we need to obtain the derivatives of the gravitational potentials.
A detailed description of the method we employ to compute the potentials and their derivatives for all the matter components of the Galaxy can be found in Appendix \ref{app:poisson}.
The derivative of the total Milky Way potential, split in all its matter components, is given by
\begin{alignat}{3}
\derp{\Phi_\mathrm{tot}}{x_i}(\tvec x) =
	& \ \derp{\Phi_\mathrm{DM}}{x_i}\left(r, \rho_\mathrm{DM}, R_\mathrm{DM}, R_\mathrm{vir,DM}\right)
	&& \quad\textrm{dark matter (eq.\ \ref{eq:nfw_profile}/\ref{eq:einasto_profile})}
	\nonumber \\
    & +\derp{\Phi_\mathrm{b}}{x_i}\left(r, \rho_\mathrm{b}, R_\mathrm{b}\right)
    && \quad\textrm{stellar bulge (eq.\ \ref{eq:bulge_rho})}
   	\nonumber \\
	& +\derp{\Phi_\mathrm{d}}{x_i}\left(R, z, \rho_\mathrm{d}, R_\mathrm{d}, z_\mathrm{d} \right)
	&& \quad\textrm{stellar disk (eq.\ \ref{eq:exponential_disk})}
	\nonumber \\
	& + \derp{\Phi_\mathrm{w}}{x_i}\left(R, z, \rho_\mathrm{w}, R_\mathrm{w}, z_\mathrm{w} \right)
	&& \quad\textrm{warm dust (eq.\ \ref{eq:exponential_disk})}
	\nonumber \\
	& + \derp{\Phi_\mathrm{c}}{x_i}\left(R, z, \rho_\mathrm{c}, R_\mathrm{c}, z_\mathrm{c} \right)
	&& \quad\textrm{cold dust (eq.\ \ref{eq:exponential_disk})}
	\nonumber \\
	& + \derp{\Phi_\mathrm{H_2}}{x_i}\left(R, z, \rho_\mathrm{H_2}, R_\mathrm{H_2}, z_\mathrm{H_2} \right)
	&& \quad\textrm{H}_2\textrm{ (eq.\ \ref{eq:exponential_disk})}
	\nonumber \\
	& + \derp{\Phi_\mathrm{HI}}{x_i}\left(R, z, \rho_\mathrm{HI}, R_\mathrm{HI}. z_\mathrm{HI} \right)
	&& \quad\textrm{HI (eq.\ \ref{eq:exponential_disk}).}
\label{eq:MW_potential}
\end{alignat}

%======================================================================
\subsection{Other objects: Virgo \& Andromeda}
\label{sec:Virgo_Andromeda}

We also wish to incorporate in our system nearby objects whose presence may have a significant impact
on the clustering factor of neutrinos in the Milky Way.
Results from $N$-body simulations in \cite{Villaescusa+11} (see their Figure 2)
show that the neutrino halo of Virgo-like clusters may extend up to distances comparable
to the one between the Milky Way and the Virgo cluster itself.
The neutrino overdensity caused by the Virgo halo at the Milky Way distance is expected
to be of a few percent, even for the minimum masses allowed by neutrino oscillations ($\Sigma m_\nu=60$~meV).
At the location of the Earth, we thus expect the Virgo effect to be almost of the same order of magnitude as the Milky Way effect.

We assume that the dark matter halo of the Virgo cluster
follows a NFW profile with a mass of $6.9\times 10^{14} \ M_\odot$ \cite{Fouque+01}.
Its distance and position in the sky in Galactic coordinates are taken
from the NASA Extragalactic Database%
\footnote{\url{https://ned.ipac.caltech.edu/}}:
\begin{equation}
    \begin{dcases}
        D_\mathrm{Virgo}  &\approx 16.5(\pm 2.0) \ \mathrm{Mpc} \\
        \mathrm{latitude}_\mathrm{Virgo}  &= 74.44^\circ \\
        \mathrm{longitude}_\mathrm{Virgo} &= 283.81^\circ
    \end{dcases}
\xRightarrow{}
    \begin{dcases}
        x_\mathrm{Virgo} = 1.056 \ \mathrm{Mpc} \\
        y_\mathrm{Virgo} = -4.299 \ \mathrm{Mpc} \\
        z_\mathrm{Virgo} = 15.895 \ \mathrm{Mpc}
    \end{dcases}.
\end{equation}

We also include in our work the Andromeda galaxy,
which is much lighter than Virgo (by a factor $\sim 500$)
but much closer (by a factor $\sim 20$) to the Milky Way.
Also for Andromeda we consider a NFW profile, but we neglect its baryon content.
The galactic latitude and longitude of Andromeda are taken from the Vizier database\footnote{\url{http://vizier.u-strasbg.fr/viz-bin/VizieR-S?NGC\%20224}},
while its distance, mass and density profile parameters are given by \cite{Kafle+18}, leading to
\begin{equation}
    \begin{dcases}
        D_\mathrm{And}  &\approx 0.784\pm 0.120 \ \mathrm{Mpc} \\
        \mathrm{latitude}_\mathrm{And}  &= -21.573311^\circ \\
        \mathrm{longitude}_\mathrm{And} &= 121.174322^\circ
    \end{dcases}
\xRightarrow{}
    \begin{dcases}
        x_\mathrm{And} = -0.377 \ \mathrm{Mpc} \\
        y_\mathrm{And} = 0.623 \ \mathrm{Mpc} \\
        z_\mathrm{And} = -0.288 \ \mathrm{Mpc}
    \end{dcases}.
\end{equation}

The density parameters at $z=0$ are listed in Table \ref{tab:dm_profile_parameters_z0_Virgo_And} for both Andromeda and the Virgo cluster.
The evolution of the density profile parameters for these objects is governed
by the same equations as for the Milky Way halo (see Section \ref{sec:mw_potential}).

\begin{figure}
\centering
\includegraphics[width=.6\linewidth,clip,viewport=500 100 1500 1300]{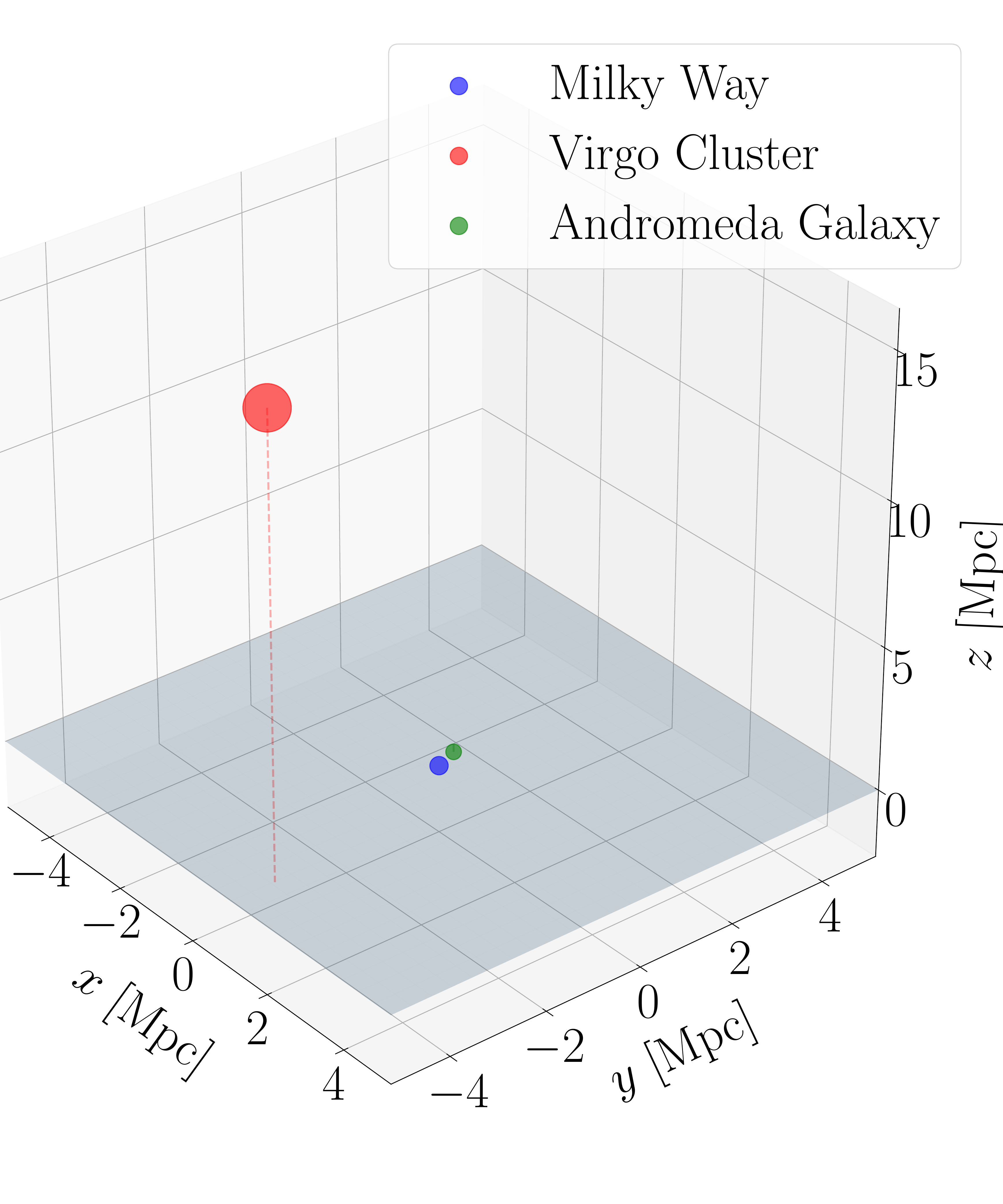}
\caption{Relative position of the Milky Way, Andromeda Galaxy and the Virgo Cluster.
The size of the dots matches the virial radius of the object.
The grey shaded plane represents the plane of the Milky Way.}
\label{fig:MW_Virgo_Andromeda_position}
\end{figure}

The complete astrophysical setup we consider, with the Milky Way, Andromeda and the Virgo cluster,
is shown in Figure \ref{fig:MW_Virgo_Andromeda_position}.
The size of the dots corresponds to the virial radius of the NFW halos.
We can note that Virgo is much bigger and distant than Andromeda.
However, as $N$-body simulations show \cite{Villaescusa+11},
the neutrino halo of each object is always much more extended than the one of dark matter,
due to the high neutrino thermal velocities.
Thus, despite its high distance from the Milky Way,
we expect that Virgo will also contribute
to the neutrino overdensity at the Earth location.

\begin{table}
\centering
\begin{tabular}{c|cc}
     & \textbf{Virgo cluster} & \textbf{Andromeda} \\
  \hline
  \hline
  $M_\mathrm{vir}$ [$M_\odot$] & $6.9\times 10^{14}$ & $8.00\times 10^{11}$ \\
  \hline
  $\rho_0$ [$M_\odot$/kpc$^3$]     & $8.08\times 10^5$   & $3.89\times 10^6$ \\
  \hline
  $R_s$ [kpc]                  & 399.1               & 21.8 \\
  \hline
  $R_\mathrm{vir}$ [kpc]       & 2328.8              & 244.7 \\
  \hline
\end{tabular}
\caption{
Dark matter density parameters for the Andromeda galaxy and the Virgo cluster at $z=0$.
The parameters for Virgo are taken from \protect\cite{Fouque+01},
and for Andromeda from \protect\cite{Kafle+18}.
}
\label{tab:dm_profile_parameters_z0_Virgo_And}
\end{table}

%======================================================================
\subsection{Gravitational potential grid}
\label{sec:grid}

Solving the Hamiltonian equations of motion requires the derivative of the gravitational potentials listed in eq.~\eqref{eq:MW_potential}.
It is convenient to provide these derivatives explicitly to the code in order to benefit from the use of textures in the GPU calculations.

First of all, we safely assume that outside the virial radius of each dark matter halo,
the potential is just given by Kepler's formula.
In this way we do not have to build very broad grids.

Inside the halos, the choice of the grid size depends on how much we want to characterize the halo itself.
For us, the most interesting structure is of course the Milky Way.
We want our grid to be much finer than the distance between the Earth and the Galactic center ($\approx$ 8 kpc)
in order to follow very accurately the trajectories of neutrinos
in the regions surrounding the Earth.
At the same time, the grid must extend at least to the maximum value (across the redshift range considered in our simulation) of the virial radius of the Milky Way, which is approximately 450 kpc.
We opt for 0.1~kpc-wide radius bins for the dark matter halo.

Likewise, for the Andromeda galaxy we also use a binning of 0.1 kpc with an extension of 350 kpc,
i.e.\ $\sim$ 50 kpc more than the maximum virial radius at $z=4$.
On the other hand, despite the fact that the Virgo cluster is much more extended than the Milky Way
(its virial radius reaches up to 3 Mpc), we do not need a very narrow binning there,
since we are not interested in what happens on very small scales.
We use a 1 kpc bin size in radius.

After computing these derivatives in spherical coordinates as a function of the radius, we get the derivatives in
Cartesian coordinates by means of the chain rule (see Appendix \ref{app:poisson}).

\begin{figure}
  \centering
  \includegraphics[width=.7\linewidth,clip,viewport=70 180 1320 1320]{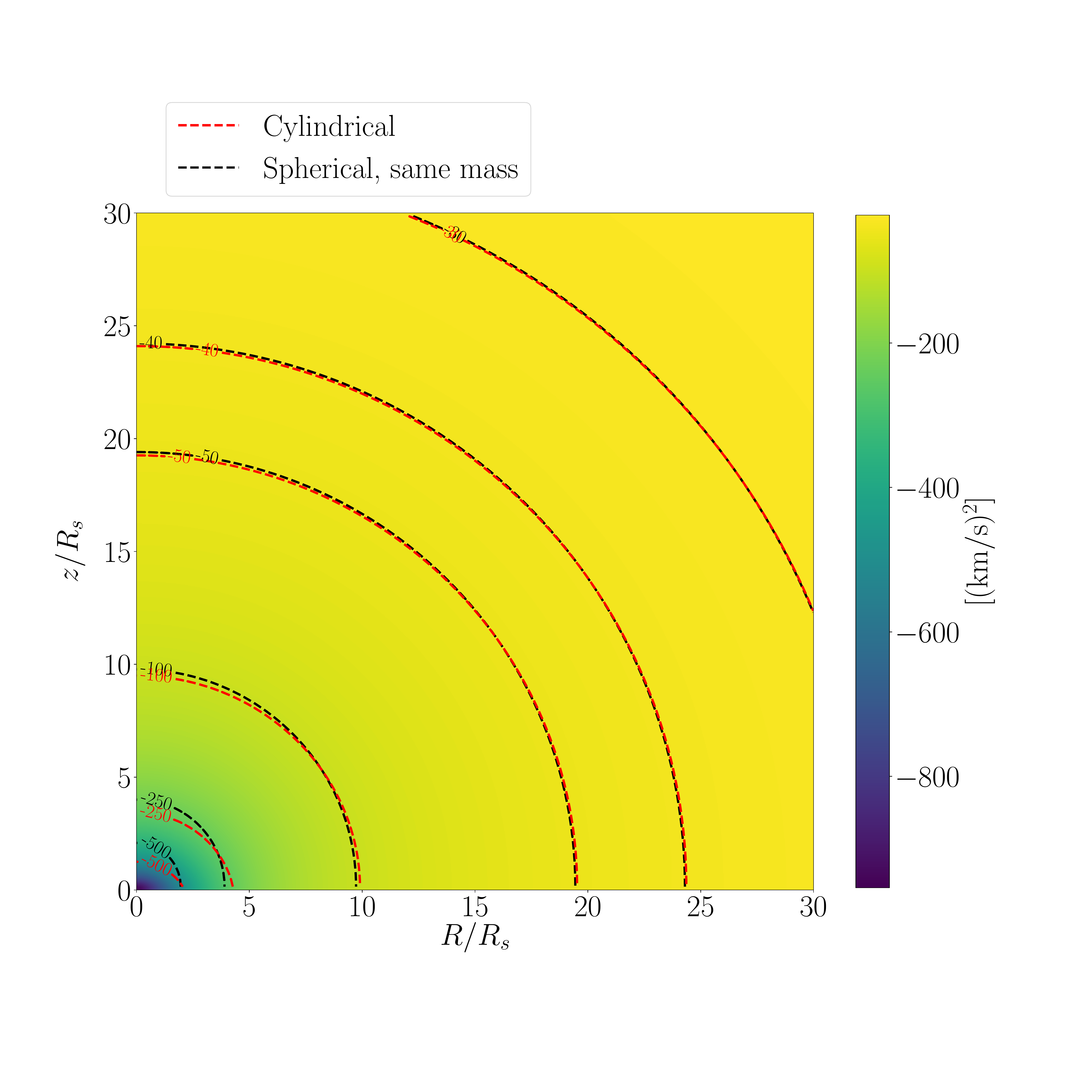}
  \label{fig:Phi_sphere}
\caption{
The colormap shows the potential generated by an exponential disk.
The red lines denote isocontours for this potential, while the black ones denote the isocontours
for the potential generated by a point-like source with the same mass.
At $R/R_s\sim z/R_s\sim25$ the difference between the spherical and cylindrical potentials is smaller than 1\%.
}
\label{fig:spherical_potential}
\end{figure}

The baryonic components only have cylindrical symmetry, leading to a more subtle situation. Their
2-D grid in $R$ and $z$ must extend at least up to a point where we can safely approximate
the potential generated by a disk-like profile with the one generated by a sphere of the same mass.
This depends of course on the ratio of scale radius and scale height: the larger the ratio, the further the grid needs to extend before we approach a Keplerian law.
Looking at Table \ref{tab:baryon_profile_parameters_z0}, we see that in the Milky Way
the maximum ratio between the scale radius and the height of the disk is 50 (for cold dust).
For this configuration, we compute the potential of an exponential profile as well as its Keplerian counterpart
(i.e.\ a point-like object with the same mass)
to check where the two potentials start to be equal to each other.
In Figure \ref{fig:spherical_potential}, we plot the different iso-potential lines for these two configurations:
it is easy to see that at distances of $R\approx \ 25 R_s$, the red and black isocontours,
which refer to the cylindrical and spherical case respectively, differ approximately by just 1\%.
We therefore extend the grid
on which we calculate
the derivative of the potential
to at least 30 times the largest scale radius among all the components.
All in all, for the Milky Way, we compute the derivative of the potential
up to $\approx$ 550 kpc from the Galactic center.

The bin sizes must be chosen carefully, especially along the direction $z$ orthogonal to the baryonic disks.
For dark matter, a bin size of 0.1~kpc would be sufficient, but some of the baryonic components have a disk much thinner than that.
We therefore opt for a logarithmic grid in $z$ that spans from $10^{-4}$ to 550 kpc.

All the above choices are
summarized in Table \ref{tab:grid_potential}.

\begin{table}
\renewcommand{\arraystretch}{1.2}
\centering
\begin{tabular}{c||c|c|c|}
&
\textbf{Milky Way} &
\textbf{Andromeda} &
\textbf{Virgo} \\
\hline
$r$ / $R$                 & $0.1-550$ kpc     & $0.1-350$ kpc & $1-3000$ kpc \\	
\hline
$\Delta r$ / $\Delta R$   & 0.1 kpc           & 0.1 kpc       &  1 kpc \\
\hline
$z$                       & $10^{-4}-550$ kpc &               &  \\	
\hline
$\Delta\log_{10}(z)$      & 0.0337            &               &
\end{tabular}
\caption{Characteristics of the grid used for the derivative of the contributions to the potential.}
\label{tab:grid_potential}
\end{table}

%======================================================================
%======================================================================
\section{Results}
\label{sec:results}

\begin{figure*}
\centering
\includegraphics[scale=1.07,clip,viewport=26 17 320 272]{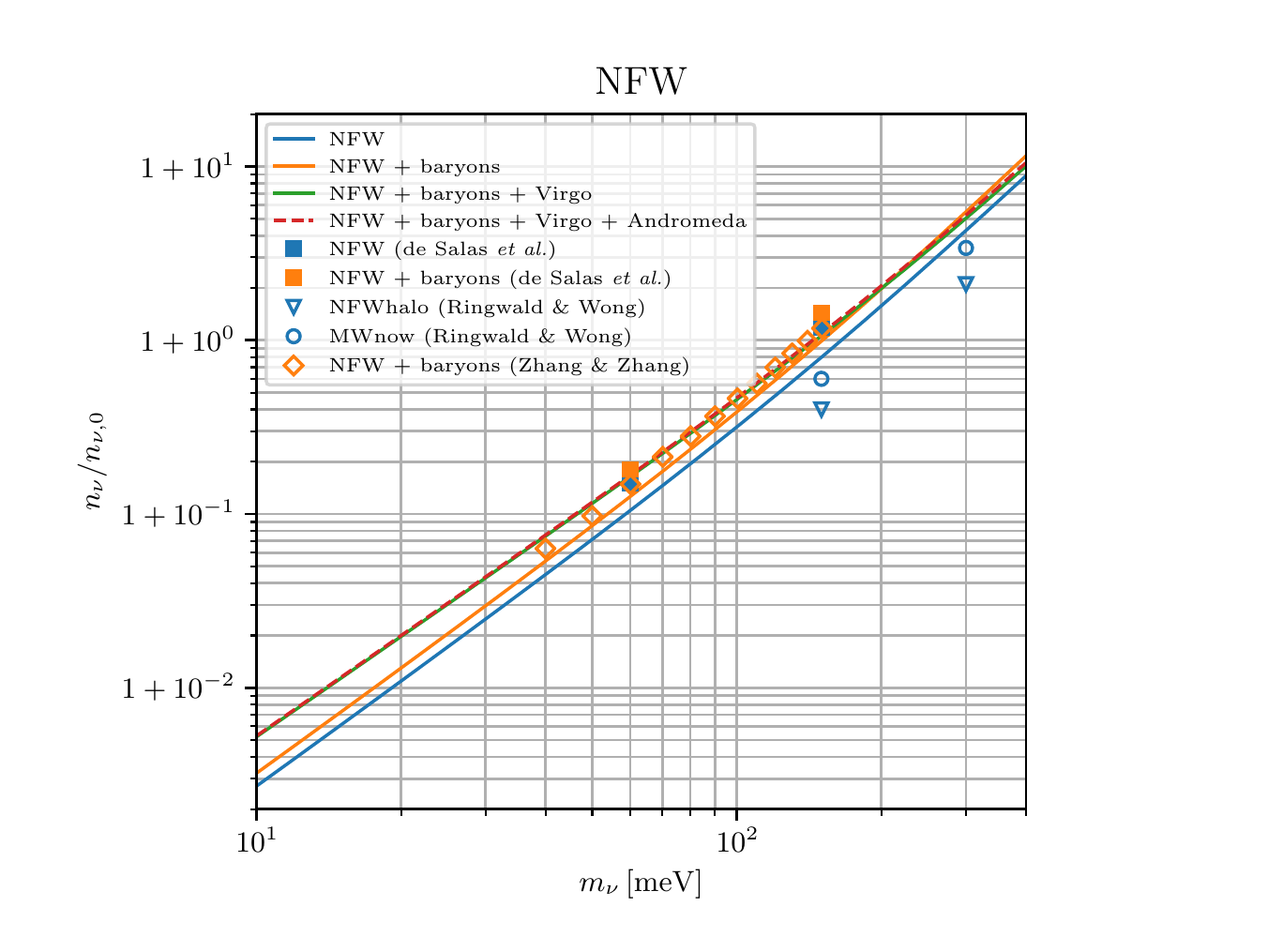}
\\[0.4cm]
\includegraphics[scale=1.07,clip,viewport=26 17 320 272]{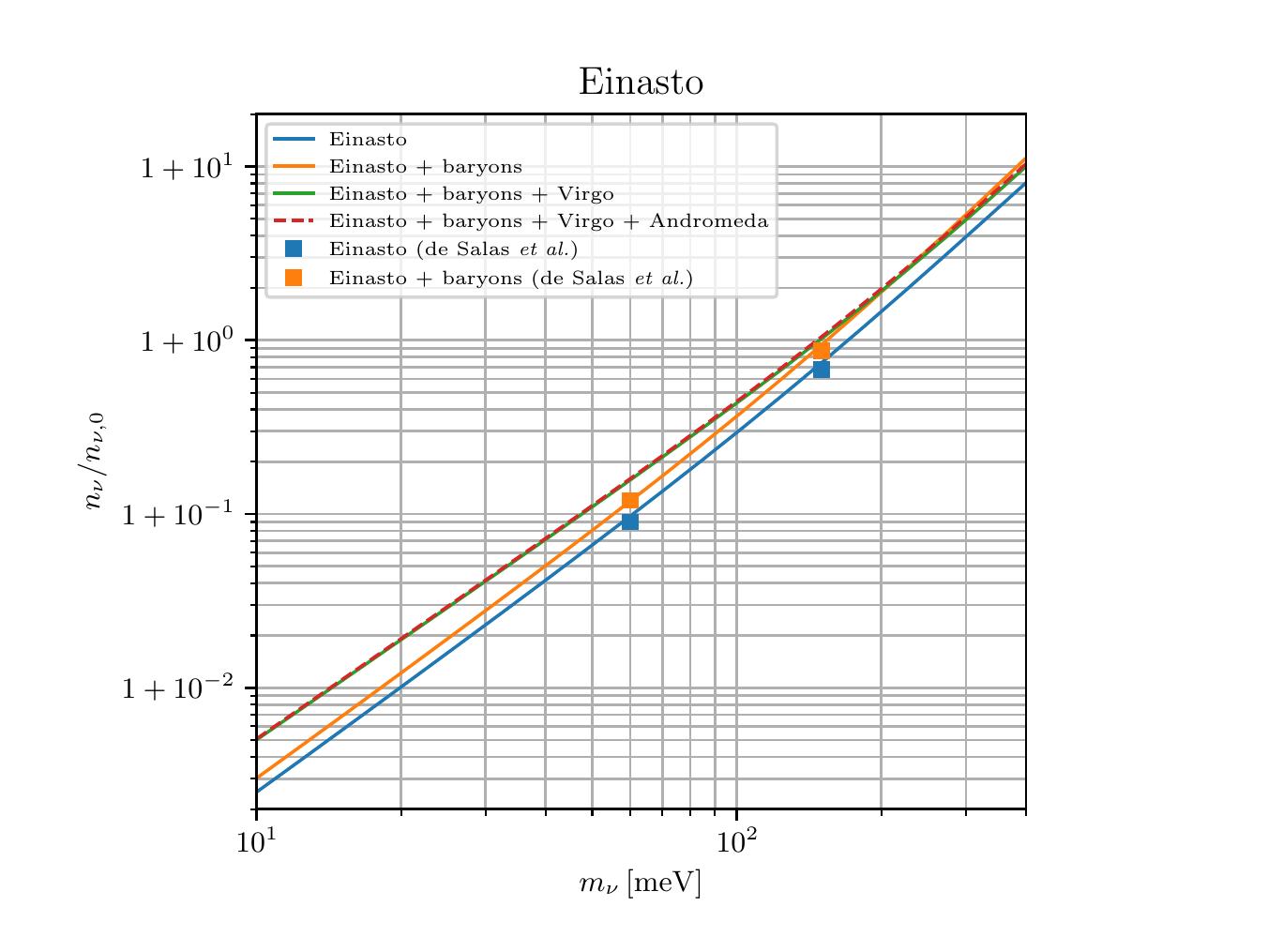}
\caption{For each neutrino mass state, we plot the ratio $n_{\nu}/n_{\nu,0}$ at the Earth's position as a function of the neutrino mass $m_{\nu}$.
We consider contributions to the gravitational potential from the Galactic dark matter halo
(\textit{top panel:} NFW profile, \textit{bottom panel:} Einasto profile), from baryons in the Galaxy, from the Virgo cluster and from the Andromeda galaxy.
We also compare with earlier studies~\protect\cite{Ringwald:2004np,deSalas:2017wtt,Zhang:2017ljh}.
}
\label{fig:results_overdensity}
\end{figure*}

In Figure~\ref{fig:results_overdensity}, we show the clustering factor, i.e.\ $n_{\nu_i}/n_{\nu,0}$,
at the Earth's position for a given neutrino mass eigenstate as a function of $m_{\nu_i}$, both for the case with an NFW distribution and an Einasto distribution for the dark matter in the Milky Way.
For Virgo and Andromeda, we only consider dark matter with an NFW profile.
We also compare our results with those of previous studies~\cite{Ringwald:2004np,deSalas:2017wtt,Zhang:2017ljh}.

As expected, regardless of our assumptions on the gravitational potential, the clustering factor increases with the neutrino mass.
The impact of baryons in our Galaxy is significant for any value of the mass.
In contrast, adding the Virgo contribution leads to an enhancement at small neutrino mass, but can actually lead to less clustering at masses larger than approximately 200~meV.
In the forward-tracking picture, this is easily explained as some of the neutrinos that would have clustered at the Earth's position
in the absence of Virgo are now clustering in the Virgo potential well instead.
In the back-tracking picture, a fraction of the particles sent out from the Earth that would have lost energy
by leaving the Milky Way's gravitational potential have fallen into Virgo's gravitational potential instead.
This leads to an increase in momentum of these particles with increasing redshift.
Thus the phase-space density is sampled only at large momenta for these particles (instead of all momenta), and the clustering is overall less pronounced.

We can also see in Figure~\ref{fig:results_overdensity} that both the effect of Andromeda and the difference between an NFW and an Einasto profile for the Milky Way's dark matter are negligible.
Assuming $m_\nu = 50 \, \text{meV}$, the overdensity is
$(n_{\nu}/n_{\nu,0} - 1) \simeq 7 \, \%$, $9 \, \%$ and $12 \, \%$ for the cases with
dark matter only, dark matter + baryons and dark matter + baryons + Virgo.

Our results are overall consistent with previous studies.
Our clustering factor is significantly larger than that inferred by \cite{Ringwald:2004np}, but with a similar dependence on the neutrino mass.
The larger clustering is likely due to our updated dark matter profile parameters.
Our results are even closer to those of \cite{deSalas:2017wtt,Zhang:2017ljh}, although slightly smaller
in the NFW case (due to different assumptions on the NFW parameters), both for the dark matter only case and for the case with baryonic contributions.

\begin{figure*}
\centering
\includegraphics[scale=1.,clip,viewport=5 25 340 510]{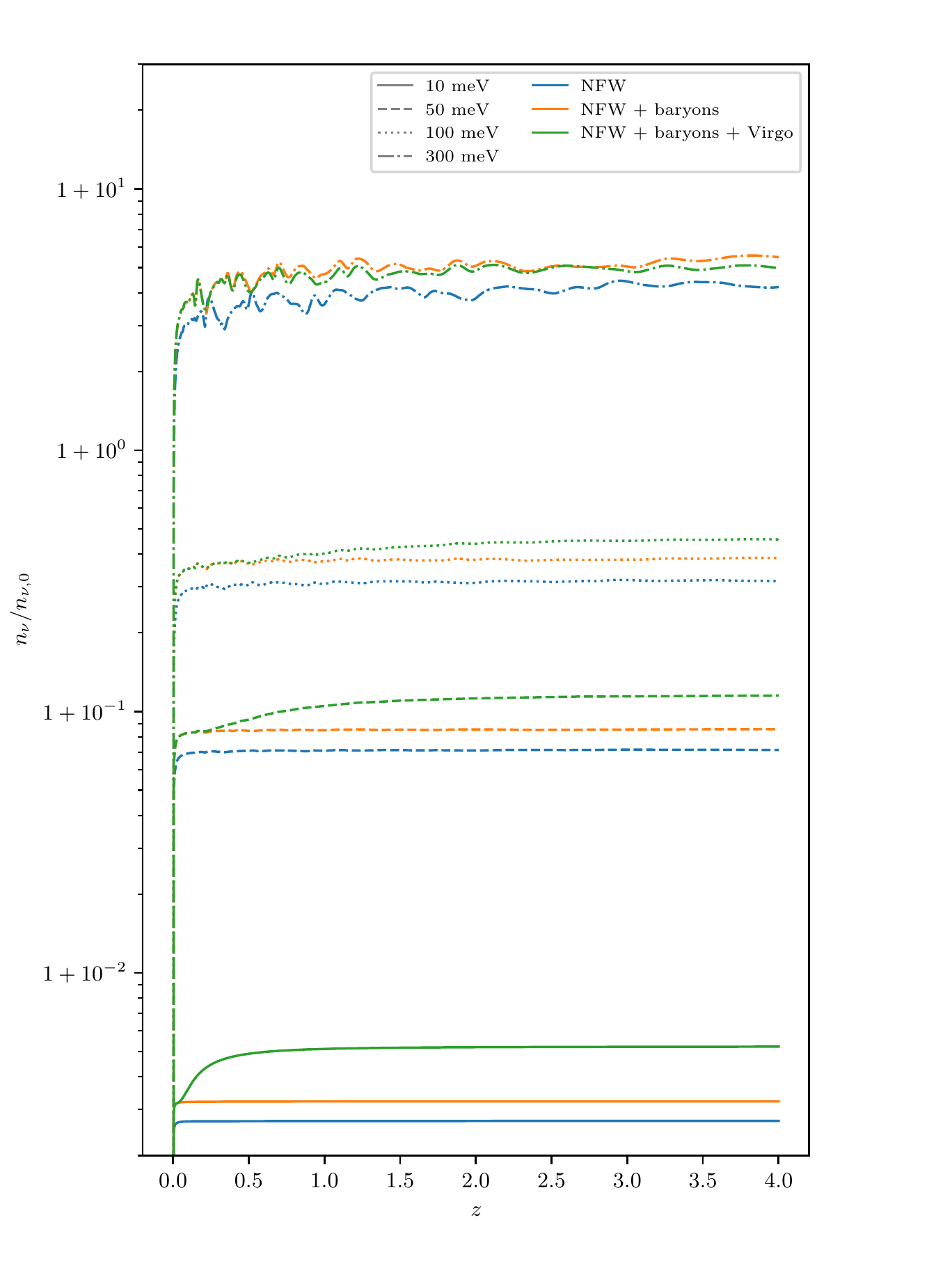}
\caption{Clustering factor as a function of the earliest redshift $z_{\rm back}$ at which neutrino trajectories are integrated,
for different values of the neutrino mass and different astrophysical configurations.
}
\label{fig:clustering_z}
\end{figure*}

Finally, we have tested the convergence of our results as a function of the redshift $z_{\rm back}$.
In the back-tracking approach, $z_{\rm back}$ controls the time at which we stop integrating the neutrino trajectories -- while with forward-tracking it would be the {\it initial} redshift.
In both cases, $z_{\rm back}$ gives the time at which we assume a perfect homogeneous and isotropic Fermi-Dirac distribution for the neutrinos.
Figure~\ref{fig:clustering_z} shows the reconstructed value of the clustering factor today when $z_{\rm back}$ is floated -- rather than being fixed to our baseline 
case of averaging $z_{\rm back} \in [3.5 , 4]$.

Figure~\ref{fig:clustering_z} shows a strong variation of the clustering factor when $z_{\rm back}$ is floated in the range from 0 to 0.5, and a gradual convergence towards an asymptotic value for $z_{\rm back}>1$.
This shows that most of the neutrino clustering takes place at very small redshift. This check is crucial for at least two reasons.
First, it shows that our slightly over-simplistic assumptions concerning the evolution of the dark matter and baryon density profiles at very high redshift cannot affect the results significantly: what matters most is to capture the gravitational potential behavior at $z<0.5$.
Second, this convergence test proves that it is sufficient to assume a perfect homogeneous and isotropic Fermi-Dirac distribution for the neutrinos at $z_{\rm back}$.
Indeed, in principle, one should either push the simulation up to $z_{\rm back} \rightarrow \infty$, or introduce some small phase-space density fluctuations $\delta f(t_{\rm back},\vec{x},\vec{p})$ accounting for the amount of clustering that took place between the onset of structure formation and $z_{\rm back}$.
If gravitational potential wells at $z_{\rm back}$ were so large that such fluctuations should be taken into account, neglecting them would introduce a bias in the results, that would depend on $z_{\rm back}$.
A non-observation of this dependence shows that the clustering between $z \rightarrow \infty$ and $z_{\rm back}$ can be safely neglected.

As one can see from Figure~\ref{fig:clustering_z},
for masses below 100~meV, the convergence of the clustering factor is achieved for $z_{\rm back}>2$.
Instead, when the neutrino mass grows,
we note that the solution is slightly less converged,
due to the existence of trapped orbits for some of the neutrinos around the Milky Way and Virgo halos,
which originate%
\footnote{In the forward picture, it is easier to understand the phenomenon:
since neutrinos are already clustering around the Milky Way and the Virgo cluster at $z=4$,
their momentum distribution function is not the homogeneous and isotropic Fermi-Dirac at such redshifts.
In the backward case, one has to think that the neutrinos cannot escape the Milky Way and the Virgo cluster
until higher redshifts.}
well before $z=4$.
In these cases, the value of $z_{\rm back}$ can have an impact on the results, but the magnitude of the oscillations seen in Figure~\ref{fig:clustering_z} shows that this is at most a 10\% effect for $n_{\nu}/n_{\nu,0}-1$.
Since this effect is smaller than the uncertainties coming from the assumptions
on the dark matter and baryon composition of the Galaxy,
and that neutrino masses above 100~meV are disfavored by cosmological measurements,
we simply present the results (Figure~\ref{fig:results_overdensity})
at high masses as an average of the values $n_{\nu}/n_{\nu,0}$ obtained considering
$z_{\rm back} \in [3.5 , 4]$.

%======================================================================
%======================================================================
\section{Conclusions}
\label{sec:conclusions}
The cosmic neutrino background, one of the predictions of the standard cosmological model,
has never been directly detected.
Due to their extremely small energy,
this is remarkably challenging, although
there are ongoing efforts to build the first relic neutrino detector.
In particular, the PTOLEMY project \cite{Baracchini:2018wwj},
under development at Gran Sasso National Laboratory in Italy,
will use the capture of relic neutrinos by tritium \cite{Weinberg:1962zza,Cocco:2007za}
as the process meant to unveil the existence of such elusive particles \cite{Betti:2019ouf}.
The event rate for the neutrino capture process depends on the density of relic neutrinos
at the location of the Earth,
which is expected to be larger than the average cosmological density ($n_0\simeq56$~cm$^{-3}$)
due to the gravitational attraction of our Galaxy.

Past works studied the clustering of relic neutrinos in the Milky Way
using a method called ``$N$-one-body simulations'' \cite{Ringwald:2004np,deSalas:2017wtt,Zhang:2017ljh},
which consists in computing the trajectories of $N$ independent test particles,
sampled assuming homogeneous and isotropic initial conditions at high redshift,
and in using the obtained information to reconstruct the shape of the neutrino halo of the Galaxy.
This method was found to be tractable only when assuming a spherically symmetric distribution of matter in our Galaxy and around: relaxing this assumption
would require an enormous number of trajectory calculations
to properly sample the six-dimensional initial phase space.
In the context of the propagation of cosmic rays in the magnetic field of the Galaxy, a similar problem
has been addressed since many years with a back-tracking method:
instead of evolving the trajectories from all the possible initial positions and momenta,
one can compute them backwards from the location of the Earth. In that way, one only considers the trajectories that are relevant for the calculation of the local cosmic rays flux.

In this paper, we used the back-tracking technique to expand the reach of the original $N$-one-body method
beyond the spherically symmetric case.
We included in our calculation a more realistic (cylindrical) description of the baryonic components
of our Galaxy,
as well as the contribution of two close-by astrophysical objects:
the Andromeda galaxy and the Virgo cluster.
We found that the main contribution comes from the dark matter halo of the Milky Way,
especially for the largest considered neutrino masses.
However, the Virgo cluster must be taken into account in order to obtain the correct number density
for the smallest neutrino masses.
The effect of the Virgo cluster is not trivial,
as its presence may actually divert some of the neutrinos that would otherwise cluster on the Milky Way
if their mass (velocity) was large (small) enough.
On the other hand, we find that the nearest galaxy with a reasonable size, Andromeda,
is giving an almost negligible contribution to the overall clustering.
For this reason, other nearby galaxies can be safely ignored.

We conclude quoting our results for a few representative values of neutrino masses.
The two cases $m_\nu = 10$~meV and $50$~meV are particularly interesting, because they stand for
plausible values of the mass of the second and third neutrino mass eigenstates in the minimal normal hierarchy scenario (that is,
when the lightest neutrino is massless and the mass ordering is normal).
Additionally, in the minimal inverted hierarchy scenario (when the lightest neutrino is massless and the mass ordering is inverted), the two heaviest neutrinos have a mass close to $m_\nu = 50$~meV.
We also quote our results for a mass of $300$~meV, in tension with recent cosmological bounds, but still well below the strong and model-independent limit currently set by KATRIN~\cite{Aker:2019uuj}.

For such masses of $10$~meV, $50$~meV, $100$~meV and $300$~meV, we obtain
that the local number density of the relic neutrinos
is respectively enhanced by 0.53\%, 12\%, $50\%$ and 500\%.
with respect to the cosmological average.
We therefore find that the local number density of relic neutrinos is
$56.8$~cm$^{-3}$, $63.4$~cm$^{-3}$, $85$~cm$^{-3}$ and $300$~cm$^{-3}$ for these cases.
% cosmological average: 339.5/6. = 56.58

%======================================================================
%============================ACKNOWLEDGEMENTS==========================
%======================================================================

\acknowledgments
SG and SP were supported by the Spanish grants SEV-2014-0398 and FPA2017-85216-P (AEI/FEDER, UE), PROMETEO/2018/165 (Generalitat Valenciana) and the Red Consolider MultiDark FPA2017-90566-REDC, as well as the
European Union's Horizon 2020 research and innovation program under the Marie Sk\l{}odowska-Curie individual Grant Agreement No.\ 796941 (SG).
They, together with GP, thank the Institute for Theoretical Particle Physics and Cosmology (TTK) at RWTH Aachen University for hospitality and support during the development of this work.
PFdS acknowledge support by the Vetenskapsr{\aa}det (Swedish Research Council) through contract No. 638-2013-8993 and the Oskar Klein Centre for Cosmoparticle Physics.
GP was supported by the INFN INDARK PD51 grant.

%======================================================================
%======================================================================
%======================================================================
%==============================APPENDICES==============================
%======================================================================
%======================================================================
%======================================================================
\appendix

%======================================================================
%======================================================================
\section{Solving the Poisson equation}
\label{app:poisson}

In order to correctly determine the clustering factor of neutrinos in the Earth neighborhood, an accurate modelling of the gravitational potential, not only of the Milky Way but also of the surrounding structures, is required.
The potential is related to the matter density via the well-known Poisson equation:
\begin{equation}
    \nabla^2\Phi(\tvec x) = 4\pi G \rho(\tvec x),
    \label{eq:poisson}
\end{equation}
where $\tvec x$ is a physical (i.e.\ not comoving) coordinate.

In case of spherical symmetry, equation \eqref{eq:poisson} is easy to solve:
\begin{equation}
\Phi(r) = -4\pi G \left[\frac{1}{r}\int_0^r \de x \ x^2 \ \rho(x) + \int_r^{\infty} \de x \ x \ \rho(x) \right].
\label{eq:poisson_spherical_symmetry}
\end{equation}
This solution can be applied to all the components of the Galaxy which satisfy spherical symmetry, namely the dark matter halo and the bulge.

The other baryonic components of our Galaxy which we wish to incorporate, such as gas and stars, are distributed in a disk, so that a more general solution to the Poisson equation must be used.
In terms of Green's functions such solution writes
\begin{equation}
    \Phi(\tvec x) = -G \int \de^3\tvec r \frac{\rho(\tvec r)}{|\tvec{x-r}|},
\end{equation}
but the integral becomes more and more difficult to solve with the increasing complexity of the system.
For this reason we employ a different approach.
When Fourier transforming both sides of equation \eqref{eq:poisson},
the Laplacian operator converts into a factor $-|\tvec k|^2$, independently of the geometry of the system.
Moving this factor at right-hand side and switching back to configuration space gives
\begin{equation}
\Phi(\tvec x) = - 4\pi G \int\frac{\de^3\tvec k}{(2\pi)^3}  \ \frac{e^{i\tvec{k\cdot x}}}{|\tvec k|^2} \ \left[\int \de^3\tvec r\  \rho(\tvec r) \ e^{-i\tvec{k\cdot r}} \right].
\label{eq:poisson_solution_with_Fourier}
\end{equation}
This is the equation we are going to use for the baryon components in the Milky Way.

%----------------------------------------------------------------------
\subsection{Spherical symmetry: the dark matter halo}

For dark matter halos, we consider two distinct cases of density profiles,
namely the NFW and Einasto profiles,
written in equations \eqref{eq:nfw_profile} and \eqref{eq:einasto_profile}, respectively.
We solve the Poisson equation
by simply working out the integrals in equations \eqref{eq:poisson_spherical_symmetry}:
\begin{align}
    \Phi_\mathrm{NFW}(r) &= -4\pi G \rho_0 R_s^2 \left[ \frac{\ln\left(1+\frac{m}{R_s}\right)}{r/R_s} - \frac{R_\mathrm{vir}/M}{1+\frac{R_\mathrm{vir}}{R_s}}\right]
    \label{eq:nfw_potential}\\
    \Phi_\mathrm{Ein}(r) &= -\frac{4\pi G \rho_0 R_s^2}{\alpha} \times
    \nonumber \\
    & \times
    \left[Y^{-1/\alpha} \ \Gamma\left(\frac{3}{\alpha},0,Y\right)+ \Gamma\left(\frac{2}{\alpha},Y,\infty\right)\right],
    \label{eq:einasto_potential}
\end{align}
where $m=\mathrm{min}(r,R_\mathrm{vir})$,
$M=\mathrm{max}(r,R_\mathrm{vir})$,
$Y = (r/R_s)^\alpha$
and we have defined the incomplete $\Gamma$ function:
\begin{equation}
\Gamma(a,L,U) = \int_L^U \de t \ t^{a-1} e^{-t}.
\end{equation}

To compute the derivative with respect to any axis $x_i$ we use the chain rule:
\begin{equation}
\derp{\Phi}{x_i} = \der{\Phi}{r}\derp{r}{x_i} = \der{\Phi}{r}\frac{x_i}{r}.
\end{equation}

%----------------------------------------------------------------------
\subsection{Spherical symmetry: the bulge}

As we mentioned in the paper, the bulge of the Milky Way is ellipsoidal rather than spherical,
with a ratio of semi-axes of $\sim 0.6$.
However, for our purposes we can safely neglect the difference between the semi-axes,
and consider the density profile given in equation \eqref{eq:bulge_rho}.
Again, the gravitational potential can be computed using equation \eqref{eq:poisson_spherical_symmetry}:
\begin{equation}
    \Phi_\mathrm{DeVac}(r) = -\frac{G\pi\rho_0 R_s^2}{A^{9/2}} \left[F_\mathrm{in}(X) + F_\mathrm{out}(X)\right],
    \label{eq:bulge_potential}
\end{equation}
where $X=A\left(\frac{r}{R_s}\right)^{1/4}$ and we have defined
\begin{align}
F_\mathrm{in}(X) &=
\frac{1}{16 X^4}\Bigg\{2027025 \sqrt{\pi} \ \mathrm{erf}\left(X^{1/2}\right) - 2 e^{-X} X^{1/2}\times
\nonumber \\
&
\times
\Big(128 X^7 + 960 X^6 + 6240 X^5 + 34320 X^4 + \nonumber \\
&
+ 154440 X^3 + 540540 X^2 + 1351350 X + 2027025\Big)
\Bigg\}
\label{eq:mass_in_bulge}
\\
F_\mathrm{out}(X) &=
\Bigg\{105 \sqrt{\pi} \ \mathrm{erfc}\left(X^{1/2}\right) + \nonumber \\
&
+ 2 e^{-X} X^{1/2}\left(8X^3+28X^2+70X+105\right)\Bigg\}.
\label{eq:mass_out_bulge}
\end{align}

Again, computing the derivatives of the potential \eqref{eq:bulge_potential}
is quite straightforward using the chain rule:
\begin{equation}
    \derp{\Phi_\mathrm{DeVac}}{x_i} = \der{\Phi_\mathrm{DeVac}}{X}\der{X}{r}\derp{r}{x_i} = \der{\Phi_\mathrm{DeVac}}{X} \frac{X}{4r}\frac{x_i}{r}.
\end{equation}

%----------------------------------------------------------------------
\subsection{Cylindrical symmetry: exponential profile}

Not all the components in a halo have spherical symmetry:
the dust, gas and stellar disk of the Milky Way
have an axial symmetry with respect to the Galactic plane.
We can therefore exploit cylindrical coordinates.
We assume that all the matter component which satisfy axial symmetry display an exponential profile,
given by equation \eqref{eq:exponential_disk}.

In this particular case the solution of the Poisson equation given by eq.~\eqref{eq:poisson_solution_with_Fourier} is (almost) analytical:
\begin{align}
\Phi_{\exp}(R,z) &=
-4\pi G\rho_0z_sR_s^2 \times
\nonumber \\
&\times
\int_0^\infty \de k \ k \ \frac{\frac{1}{k}e^{-k|z|} - z_se^{-|z|/z_s}}{\left[1+\left(kR_s\right)^2\right]^{3/2}\left[1-k^2z_s^2\right]} \ J_0(kR).
\label{eq:Poisson_solution_cylindrical_symmetry_exponential_profile}
\end{align}

The integral appearing in equation \eqref{eq:Poisson_solution_cylindrical_symmetry_exponential_profile}
is a Hankel transform of order 0:
we solve it using the FFTlog code%
\footnote{
Specifically we use the \texttt{python} package \texttt{FFTLog} (\url{https://github.com/prisae/fftlog})
by D.~Werthm\"{u}ller,
based on the \texttt{Fortran} \texttt{FFTLog} code by A.~Hamilton \cite{Talman:1978,Hamilton:1999uv}.
}.

\begin{figure}
\centering
\includegraphics[width=.7\linewidth]{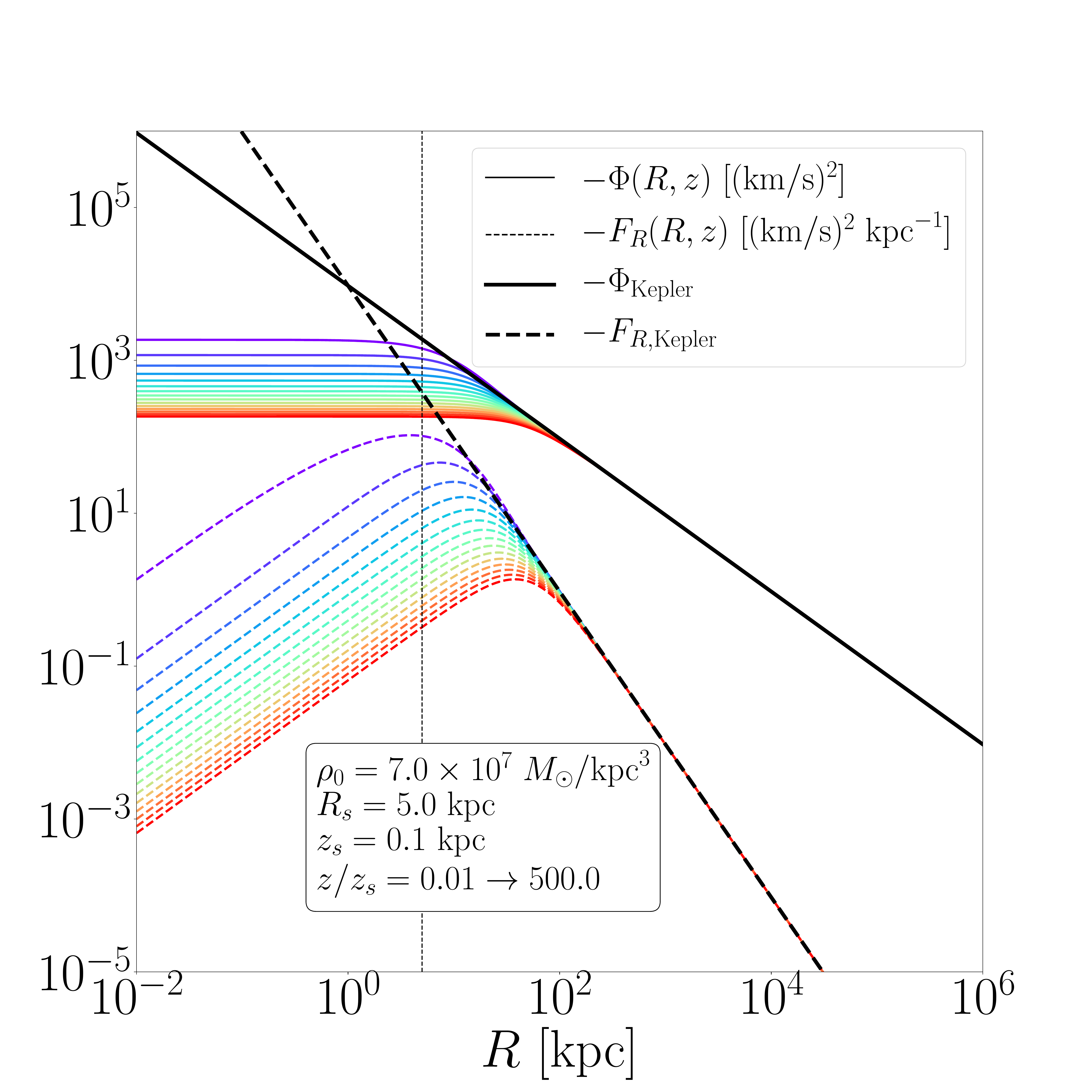}
\caption{
The colored solid lines represent the potential generated by an exponential disk
with the same parameters as the cold dust profile in the Milky Way.
Dashed lines represent the derivative with respect to $R$ of this potential,
i.e.\ the force per unit mass in the radial direction.
The different color label different $z$:
the blue curves are the closest to the Galactic plane, the red ones are the farthest ones.
For comparison, also the Keplerian potential and force are plotted (solid and dashed black lines respectively):
for radii much larger than $R_s$ (dotted vertical line)
both the potential and its derivative converge to this value.
}
\label{fig:disk_phi_force}
\end{figure}

The derivatives of this potential are also analytical.
For the $x$ and $y$ directions, we exploit the chain rule combined
with the property of Bessel functions $\der{J_\alpha(x)}{x} = -J_{\alpha+1}(x)$,
while the derivative with respect to the $z$-axis is obtained by just deriving the integrand function:
\begin{align}
\derp{\Phi_{\exp}}{(x,y)} &=
4\pi G\rho_0z_sR_s^2 \ \frac{(x,y)}{R} \times
\nonumber \\
&\times
\int_0^\infty \de k \ k \ \frac{k\left(\frac{1}{k}e^{-k|z|} - z_se^{-|z|/z_s}\right)}{\left[1+\left(kR_s\right)^2\right]^{3/2}\left[1-k^2z_s^2\right]} \ J_1(kR),
\label{eq:derphi_xy_cyl}
\\
\derp{\Phi_{\exp}}{z} &=
4\pi G\rho_0z_sR_s^2 \frac{z}{|z|} \times
\nonumber \\
&\times \int_0^\infty \de k \ k \ \frac{e^{-k|z|} - e^{-|z|/z_s}}{\left[1+\left(kR_s\right)^2\right]^{3/2}\left[1-k^2z_s^2\right]} \ J_0(kR).
\label{eq:derphi_z_cyl}
\end{align}	

The behavior of these last equations can be seen in Figure \ref{fig:disk_phi_force},
where the example of the cold dust disk in the Milky Way is shown.
The solid lines represent the gravitational potential as a function of radius for different values of $z$,
indicating with blue (red) the closest (farthest) to the Galactic plane.
Dashed lines in turn are the derivative of this potential with respect to $R$.
As can be noticed, both the potential and its derivative converge to the Keplerian limit
(solid and dashed black) when going sufficiently far outside the scale radius (dotted vertical line).

%============================ REFERENCES ==============================
\bibliographystyle{JHEP}%{unsrt}
\bibliography{bibtex}

\end{document}